\documentclass[aps,prl,floatfix,superscriptaddress,twocolumn]{revtex4}
\usepackage{graphics}
\begin{document}

\title{A robust one-step catalytic machine
for high fidelity anti-cloning and \\
W-state generation in a multi-qubit system}
\author{Alexandra Olaya-Castro}\email{a.olaya@physics.ox.ac.uk}
\author{Neil F. Johnson}
\affiliation{Centre for Quantum Computation and Department of Physics,
University of Oxford, Clarendon Laboratory, Parks Road, OX1 3PU, United Kingdom}
\author{Luis Quiroga}
\affiliation{Departamento de F\'{\i}sica, Universidad de Los Andes, A.A. 4976,
Bogot\'a , Colombia}


\begin{abstract} 
We propose a physically realizable machine
which can either generate multiparticle W-like
states, or implement high fidelity $1 \rightarrow M$ ($M=1,2,\cdots \infty$) anti-cloning
of an arbitrary qubit state, in a single step.
Moreover this universal machine acts as a catalyst in that it is unchanged after
either procedure,
effectively resetting itself for its next operation.
It also possesses an inherent {\em immunity} to decoherence. 
Most importantly in terms
of practical multi-party quantum communication, the machine's robustness in the
presence of decoherence actually {\em increases} as the number of qubits $M$ increases.
\end{abstract}

\maketitle
The mathematical foundations of quantum mechanics yield two remarkable
consequences in terms of what is possible in our Universe, and what isn't.
First, the {\em linearity} of quantum mechanics implies that it is impossible to make a
perfect copy of an arbitrary quantum (qubit) state
\cite{wootters82}, no matter how ingenious the experimental scheme. Second, the {\em
unitarity} of quantum mechanics implies that there is no quantum device, no matter how
well-built, which can perfectly transform an arbitrary qubit state into its orthogonal
complement\cite{gisin99, buzek99}. Despite these fundamental `laws of Nature', we now
know that cloning and the complementing of qubits can still be carried out with
reasonably high fidelity \cite{buzek99, buzek96}. There is even the suggestion that these two
processes might actually be closely related \cite{buzek99}. In fact, recent experiments
have demonstrated that optimal
$1\rightarrow 2$ cloning (i.e. partially copying a quantum state from one qubit onto
two target qubits) and the universal-{\footnotesize NOT} of photon polarization states,
can both be performed using the same unitary transformation \cite{demartini04,
irvine04}. Indeed, it has been suggested that a combination of  copying and
complementing could lead to optimal entangling transformations \cite{buzek00}. However,
the connections  between these two quantum processes are still not well understood
either in theoretical or practical terms. In a seemingly unrelated development, 
researchers interested in building  quantum information machines have begun to
propose experimental schemes using  `always-on'
Hamiltonian interaction terms, in order to avoid the need for switching on and off
multiple quantum gates
\cite{Bose}. However such schemes invariably assume that specific two-body,
nearest-neighbor interactions can be engineered in some particular qubit geometry
(e.g. chain) despite the fact that nanostructures, for example, may have
long-range interactions due to residual electrostatic potentials. 

In this paper, we bring together these two seemingly separate lines of research by
proposing a multiqubit-cavity scheme in which the {\em same} unitary
transformation can be used to produce multiqubit W-entangled states and high
(in some cases optimal) fidelity $1\rightarrow M$ anti-cloning, where $M$ is any
arbitrary number of qubits. As a result, our work provides a
concrete  connection between copying, complementing and entangling operations.
From a practical point of view, the implementation of our scheme
offers a number of outstanding advantages and features. First, the cavity acts as a
{\em catalyst} in that its state is unchanged after either procedure -- in short, our
machine acts as its own reset button. Second, the machine has an inherent immunity to
decoherence effects. In particular, our calculations show that entangling and
anti-cloning operations become increasingly robust as the number of qubits {\em increases},
in contrast to typical quantum information schemes whose performance would deteriorate
as the number of degrees-of-freedom increases.  Third, our machine avoids the need for
carefully engineered nearest-neighbor interactions \cite{chiara04}, 
multiple cavities and/or gate operations \cite{milman03,zou03}. Moreover, our multiqubit-cavity machine could be built
using current atom- or ion-cavity technology, or  next-generation quantum-dot or
SQUID-cavity technology.

The Hamiltonian for the $M$-qubit-plus-cavity
system in the interaction picture
and rotating-wave approximation ($\hbar=1$) is 
\begin{eqnarray}H_I=\sum_{j=1}^{M}\gamma_j\{a^\dag \sigma_j^{-}+\sigma_j^{+}a\},
\label{eq:ham}
\end{eqnarray} where $\sigma_j^{+}=|1_j\rangle \langle 0_j|$,
$\sigma_j^{-}=|0_j\rangle \langle 1_j|$ with $|1_j\rangle$ and $|0_j\rangle$
being the excited and ground states of the $j$'th qubit. Here
$a^\dag$ and
$a$ are cavity-photon creation and annihilation operators while $\{\gamma_j\}$ are
the set of (in general unequal) qubit-cavity couplings.
Since $[H_I,{\cal
N}]=0$ where ${\cal N}=a^\dag a +
\sum_{i=1}^M \sigma^{+}_i \sigma^{-}_i$ is the excitation number operator,
the dynamics is separable into subspaces having a prescribed eigenvalue $N$
of ${\cal N}$. In particular, in the subspace with $N=0$ there is only one state
$|\phi_0\rangle  =  |0_1,0_2,0_3\cdots 0_M;0\rangle$ while in the $N=1$ subspace, the basis states are
\begin{eqnarray} |\phi_1\rangle & = & |1_1,0_2\cdots 0_M;0\rangle = |Q_1\rangle
\otimes |0\rangle \nonumber \\ |\phi_2\rangle & = & |0_1,1_2\cdots
0_M;0\rangle  = |Q_2\rangle\otimes|0\rangle \nonumber\\
\vdots & = &\quad \quad \quad \quad \vdots \quad \quad \quad \quad =\quad
\vdots \quad \nonumber\\ |\phi_j\rangle & = & |0_1,0_2\cdots 1_j,\cdots
0_M;0\rangle =  |Q_j\rangle\otimes|0\rangle \\ \nonumber |\phi_{M+1}\rangle & =
& |0_1,0_2\cdots 0_M,1\rangle
\end{eqnarray}
where the last label in each ket denotes the photon number in the
cavity. In this $N=1$ subspace, the system's state at time $t$ is
$|\Psi(t)\rangle=\hat U(t,0)|\Psi(0)\rangle$ where $\hat U(t,0)$ in the basis
of states $\{ |\phi_1\rangle \cdots |\phi_{M+1}\rangle \}$ becomes
\begin{widetext}
\begin{eqnarray}
\hat U(t,0)=\left (
\begin{array}{*{5}{c}} 1-2\gamma_1^2\beta & -2\gamma_1 \gamma_2 \beta &\;
\cdots \; &-2\gamma_1\gamma_M \beta & -i\gamma_1 {\rm sin}(\omega t)/\omega\\
-2\gamma_2 \gamma_1 \beta & 1-2\gamma_2^2 \beta &\; \cdots \;
&-2\gamma_2\gamma_M \beta & -i\gamma_2 {\rm sin}(\omega t)/\omega\\
\; \vdots \; & \; \vdots \; &\; \vdots \; &\; \vdots \;&\; \vdots \;\\
-2\gamma_M \gamma_1 \beta & 2\gamma_M\gamma_2 \beta &\; \cdots \;
&1-2\gamma_M^2 \beta & -i\gamma_M {\rm sin}(\omega t)/\omega\\ -i\gamma_1 {\rm
sin}(\omega t)/\omega & -i\gamma_2 {\rm sin}(\omega t)/\omega &\; \cdots \; &
-i\gamma_M {\rm sin}(\omega t)/\omega & {\rm cos}(\omega t)
\end{array}\right )\ \ \ .
\label{eq:u}
\end{eqnarray}
\end{widetext}
The effective collective Rabi frequency of the $M$ qubits is
$\omega^2=\sum_{j=1}^{M}\gamma_j^2$, and $\beta={\rm sin}^2(\omega
t/2)/\omega^2$. We now show how we build our machine using this temporal
evolution, evaluated over a specially chosen time interval.

Consider an initial product state where one of the qubits (e.g.. $j=1$) is in
a coherent superposition $|q_1(0)\rangle={\rm sin}(\theta/2)|0_1\rangle +
e^{i\alpha} {\rm cos}(\theta/2)|1_1\rangle$ with the others
unexcited:
\begin{eqnarray} |\Psi(0)\rangle & = & |q_1(0)\rangle\otimes|0_2,\cdots
0_M;0\rangle \nonumber \\& = & {\rm sin}(\theta/2)|\phi_0\rangle + e^{i\alpha}
{\rm cos} (\theta/2)|\phi_1\rangle \label{eq:state}
\end{eqnarray}
Using  Eq.(\ref{eq:u}) yields
\begin{eqnarray} |\Psi(t)\rangle={\rm sin}(\theta/2)|\phi_0\rangle +
e^{i\alpha} {\rm cos} (\theta/2)|\phi_1 (t)\rangle
\end{eqnarray} with
\begin{eqnarray} |\phi_1(t)\rangle
=\sum_{j=1}^{M}U_{j1}(t)|Q_j\rangle\otimes|0\rangle -i\frac{\gamma_1{\rm
sin}(\omega t)}{\omega}|\phi_{M+1}\rangle\ \ .
\end{eqnarray}
When $\omega t =m\pi\equiv \omega
\tau^*$  ($m$ odd) a vacuum trapping state condition is achieved:
the cavity state is unchanged overall and becomes fully separable from the
multiqubit subsystem. However, its catalytic action has induced
entanglement into the initially unentangled multiqubit subsystem. Because of
the cavity's inertness at $t=\tau^*$, we drop the cavity state notation from now
on.

Consider the following two specific examples:
(i) $\theta=0$, which will yield one-step W-state generation; (ii)
$\theta=\pi/2$, which will yield optimal quantum anti-cloning.

\vskip0.05in
\noindent (i) Using $\theta=0$ yields
\begin{eqnarray}
|\Psi(\tau^*)\rangle=(1-2\gamma_1^2/\omega^2)|Q_1\rangle-(2\gamma_1/\omega^2)
\sum_{j=2}^{M}\gamma_j
|Q_J\rangle\ \ .
\end{eqnarray}
In general, an $M$-qubit W-state cannot be generated using
identical couplings $\gamma_i \equiv \gamma$. However for
non-identical couplings, the qubit-exchange symmetry is broken
thereby allowing control over the degree of entanglement and the
final state symmetry (see Ref. \cite{olaya04} for $M=2$). Suppose
$\gamma_1 \neq \gamma_j=\gamma$ for all $j>1$, and define
$r=\gamma_1/\gamma$. The collective qubit frequency is
$\omega=\gamma (r^2+M-1)^{1/2}$ and
\begin{eqnarray} |\Psi(\tau^*)\rangle = a_1(\tau^*)|Q_1\rangle +
a(\tau^*)\sum_{j=2}^{M}|Q_j\rangle
\label{eq:trapstate}
\end{eqnarray} where
\begin{eqnarray} a_1(\tau^*)=\frac{M-1-r^2}{M-1+r^2},\quad
a(\tau^*)=\frac{-2r}{M-1+r^2}\ \ .
\end{eqnarray}
Two $W$ states of $M$ qubits can now be generated for
$a_1(\tau^*)=\pm a(\tau^*)$, yielding an optimal coupling ratio
$r_W^{\pm}=\sqrt{M} \pm 1$. Here $r_W^{\pm}$ correspond to symmetric and
antisymmetric states with respect to exchange of qubit-1 with any other.
The corresponding state is
\begin{eqnarray} |\Psi(\tau^*)\rangle=|W_{M}^{\pm}\rangle =
\frac{e^{i\pi}}{\sqrt M}\left [\pm|Q_1\rangle + \sum_{j=2}^{M}|Q_j\rangle
\right ]\ \ .
\end{eqnarray}
For $M=4$, both $r=1$ and $r=3$ produce a fully
symmetric W state. However in the many-qubit
limit
$M\rightarrow
\infty$, it is only for non-identical couplings ($r_{W}\simeq
\sqrt{M}$) that we can generate a multi-qubit W entangled state. (N.B. A W-state is
of interest for quantum information protocols since the excitation has
identical probabilities of being found on any of the qubits). We note that for $M
\geq 3$, a fully symmetric W state of $M-1$ qubits can also be obtained when
$a_1(\tau^*)=0$, yielding
$r_{W'}=\sqrt{M-1}$. The initial excitation gets transferred to, and shared among,
the remaining $M-1$ qubits.

\vskip0.05in
\noindent (ii) Using $\theta=\pi/2$ enables us to anti-clone (i.e. copy the
orthogonal complement) of the state of qubit-1 (i.e. the input qubit) to the target
qubits (i.e. the $M-1$ qubits initially in state
$|0\rangle$). Since the initial state of qubit-1 is in the equatorial plane of
the Bloch sphere (see Eq.(4)), we will call this process Phase-Covariant
Anti-Cloning (PCAC) in analogy with phase-covariant cloning (PCC) \cite{bruss00}.
An ideal anti-cloning process is defined
as \cite{gisin99}
\begin{eqnarray} |q\rangle_{a}|0\rangle_b|D\rangle_{in}\rightarrow
|q^\perp\rangle_{a}|q^\perp\rangle_b|\widetilde{D}\rangle_{out}
\label{eq:clon}
\end{eqnarray} where $|q\rangle_{a}$ is the initial state of the input qubit,
$|0\rangle_{a}$ is the initial state of a target qubit, $\langle
q|q^\perp\rangle=0$, and $|D\rangle_{in}$ and
$|\widetilde{D}\rangle_{out}$ are the input and output states of
the copying device. Ref.\cite{bruss00} showed that the optimal
fidelity for $1\rightarrow 2$ PCC
 is $\mathcal {F}^{opt}=\frac{1}{2}[1+\frac{1}{\sqrt 2}]$. Here we demonstrate
that the fidelity of our $1\rightarrow 2$ anti-cloning equals this optimal
value. We also show that there are two protocols to achieve this process, the
main difference being the final state of the input qubit. For an arbitrary number
of output qubits $M$ with asymmetric couplings, we find that the fidelity of the
anti-cloning operation is comparable to that obtained for a XY spin star
network\cite{chiara04} and reaches larger values than for the case of
identical couplings.
The state of the system is now
$|\Psi(t)\rangle=\frac{1}{\sqrt 2}\left [|\phi_0\rangle + e^{i\alpha} |\phi_1
(t)\rangle \right ]$ and the reduced density matrix of the $j'th$ qubit reads
\begin{eqnarray}
\rho_j(t) & = & \frac{1}{2}\big[(2-|U_{j1}(t)|^2)|0_j\rangle \langle 0_j| +
|U_{j1}(t)|^2|1_j \rangle \langle 1_j|\nonumber \\ & &
U_{j1}(t)(e^{-i\alpha}|0_j \rangle \langle 1_j| + e^{i\alpha}|1_j \rangle
\langle 0_j|)\big ]
\end{eqnarray}
The fidelity of copying $|\tilde q
\rangle=(1/ \sqrt 2)[|0\rangle + e^{i \mu}|1\rangle]$ to the $j$'th qubit, is
$\mathcal {F}_{j}(t)=\langle
\tilde q|\rho_j|\tilde q\rangle =\frac{1}{2}\{1+U_{j1}(t){\rm
cos}(\alpha-\mu)\}$. For a target qubit, at $t=\tau^*$,
\begin{eqnarray}
\mathcal {F}_{j>1}(\tau^*)=\frac{1}{2}\{1-2\gamma_1\gamma_j {\rm
cos}(\alpha-\mu)/\omega^2\}\\\nonumber
\end{eqnarray}
hence the fidelity is greater than $1/2$ when the
state that has been copied corresponds to the orthogonal complement of the
input state (anti-cloning), i.e. $\alpha-\mu = \pi$.
\begin{figure}
\resizebox{8cm}{!}{\includegraphics*{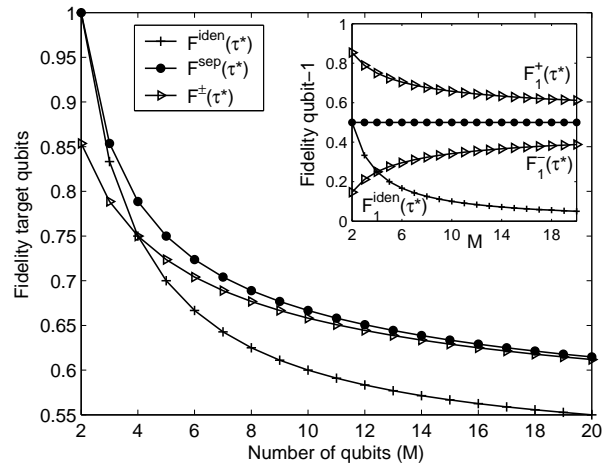}}
\caption{Anti-cloning fidelity as a function of the number of
qubits $M$. For the target qubits ($j>1$), $\mathcal{F}^{iden}(\tau^*)=\frac{1}{2}[1+~\frac{2}{M}]$ denotes
the case of identical couplings, $\mathcal
{F}^{\pm}(\tau^*)=\frac{1}{2}[1+~\frac{1}{\sqrt M}]$ denotes the
case where $r_{W}^{\pm}=\sqrt M \pm 1$ and $\mathcal {F}^{sep}(\tau^*)=
\frac{1}{2}[1+\frac{1}{\sqrt{M-1}}]$ denotes the case where
$r_{W'}=\sqrt{M-1}$. {\bf Inset}: Fidelity of original qubit
$\mathcal {F}_1$ for $r=1$ : $\mathcal {F}^{iden}_1(\tau^*)=1/M$,
$r_W^{\pm}$ : $\mathcal {F}_1^{\pm}(\tau^*)=\frac{1}{2}[1\pm \frac{1}{\sqrt M}]$,
and $r_{W'}$ : $\mathcal {F}_1^{sep}=1/2$. All results are evaluated
at the trapping time
$t=\tau^*$.}
\label{fig:fidelity}
\end{figure}
Figure 1 shows the fidelity of a target qubit as well as for the
input qubit (inset) as a function of the number of qubits. For
coupling ratio $r_W^+$, the input qubit finishes entangled with
the target qubits, i.e. $|\Psi(\tau^*)\rangle =\frac{1}{\sqrt
2}[|\phi_0\rangle + e^{i\alpha} |W_{M}^{\pm}\rangle]$ such that
the fidelity of the input qubit (see inset) equals the fidelity of
the target qubits. Hence, we obtain $M$ outputs (including the
input qubit) with fidelity $\mathcal{F}^{+}=\frac{1}{2}[
1+\frac{1}{\sqrt M}]=\mathcal {F}^+_{1}$. For $M=2$, we obtain
$\mathcal{F}^{+}_{M=2}=\frac{1}{2}[1+\frac{1}{\sqrt 2}]$ which
equals the optimal value for the $1 \rightarrow 2$ PCC
\cite{bruss00}. Interestingly, such optimal transformation
combines two operations in one-step: complementing the original
qubit's state and copying. We also note that this optimal fidelity
is achieved for the same conditions under which two-qubit
maximally entangled states were found\cite{olaya04}, hence
establishing a direct connection between optimal anti-cloning and
maximal entanglement. For $r_W^-$, the fidelity of the target
qubits equals $\mathcal{F}^{+}$ but the fidelity of the input
qubit is always less than $1/2$, i.e.
$\mathcal{F}^{-}_1=\frac{1}{2}[1-\frac{1}{\sqrt M}]$ which is
undesirable for a single qubit NOT operation. For
$r_{W'}=\sqrt{M-1}$, we obtain $M-1$ outputs with fidelity
$\mathcal {F}^{sep}=\frac{1}{2}[1+\frac{1}{\sqrt {M-1}}]$ while
the fidelity of the input qubit equals $1/2$ irrespective of the
number of qubits (see inset). This is because the input qubit ends
in its ground state and separated from the rest, i.e.
$|\Psi(\tau^*)\rangle =|0_1\rangle\otimes\frac{1}{\sqrt 2}[|\tilde
{\phi}_0\rangle + e^{i\alpha}|W_{M-1}\rangle ]$ with $|\tilde
{\phi}_0\rangle=|0_2,0_3,\dots 0_M\rangle$. For $M=3$ we obtain $1
\rightarrow 2$ anti-cloning with optimal fidelity $\mathcal
{F}^{opt}$ for the target qubits. In general, $\mathcal
{F}^+(M)=\mathcal {F}^{sep}(M+1)$ which means that there exist two
protocols for obtaining $M$ outputs with high fidelity: (i) $M$
qubits and $r=r_W^+$, and (ii) $M+1$ qubits and $r=r_{W'}$. The
main difference between these two protocols is the time operation
$\tau^*=\pi/\omega$: it is shorter for the $r_W^+$ case since
$\omega (r_W^+)>\omega(r_{W'})$. For a large number of outputs,
this difference is negligible since $r_W \simeq r_{W'}$.
Interestingly, the operation time decreases with the number of
anti-clones, implying that the protocols are robust in the
presence of decoherence. We confirm this robustness in more detail
below. In both cases $r_W^{\pm}$ and $r_{W'}$, the fidelity of the
one-step anti-cloning procedure is comparable with that reported
for cloning operations using a XY spin network \cite{chiara04}
since it depends on the number of outputs $M$ as $1/\sqrt M$. In
the case of identical couplings, the fidelity of the target qubits
is $\mathcal {F}^{iden}=\frac{1}{2}[1+\frac{2}{M}]$ which is
always less than $\mathcal {F}^{sep}$ as well as being less than
$\mathcal {F}^{\pm}$ for $M>4$. This behaviour is comparable with
that of a Heisenberg spin network since it depends on the number of
outputs $M$ as $1/M$ \cite{chiara04}.
\begin{figure}
\resizebox{8cm}{!}{\includegraphics*{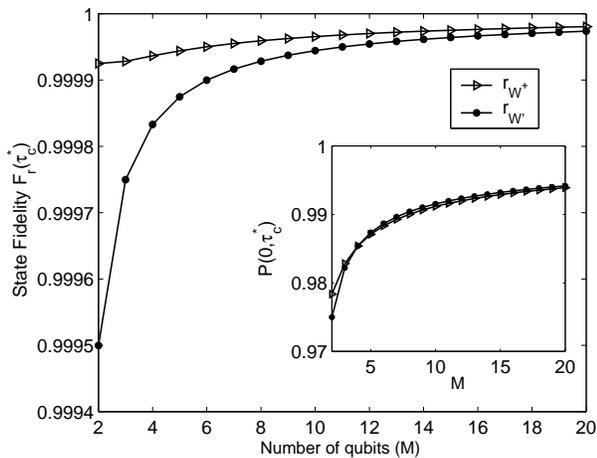}}
\caption{Fidelity of the state obtained at $\tau^*_c=2\pi/\Omega$ with respect
to the pure state obtained at  $\tau^*$ $(F_r(\tau^*_c))$ as a function of the
number of qubits $M$. Two cases are shown: $r_{W}^+$ (triangles) and
$r_{W'}$ (circles). {\bf Inset}: Probability of no photon detection during
interval $(0,
\tau^*_c)$. Here $\kappa=0.02\gamma$ and $\Gamma=0.001\gamma$.}
\label{fig:decoherence}
\end{figure}

Decoherence would take place through two main channels:  qubit dipole decay
at rate
$\Gamma$, and cavity decay with rate $\kappa$. A single trajectory in the
quantum jump model \cite{plenio} is well-suited to evaluate the effects on the
fidelity at the trapping time. We suppose that the
photon decays are continuously monitored, and that the single trajectory is
specified by the evolution of the system conditioned to no-photon detection.
The conditional dynamics satisfies the dissipative Hamiltonian
\begin{eqnarray}
\widetilde{H}=H_I -i\Gamma \sum_{j=1}^M \sigma^+_j \sigma^-_j -i\kappa a^\dag a\ \ .
\end{eqnarray}
The (unnormalized) conditional state
$|\Psi_{cond}(t)\rangle=\widetilde
U(t,0)|\Psi(0)\rangle=\sum_{j=1}^{M+1}b_{j}(t)|\phi_j\rangle$ with
$\widetilde U(t,0)={\rm exp}[-i\widetilde {H}t]$ and $\parallel
P(0,t)=|\Psi_{cond}(t)\rangle\parallel^2$ being the probability of not
detecting a photon in the interval $(0,t)$. The conditional state becomes
{\small\begin{eqnarray}
|\Psi(t)\rangle = b_1(t)|\phi_1\rangle +
b(t)\sum_{j=2}^{M}|\phi_j\rangle +b_{M+1}(t)|\phi_{M+1}\rangle
\label{eq:trapcond}\end{eqnarray}}
where
\begin{eqnarray}
b_1(t)&=&1+rb(t)\nonumber \\
b(t)&=&\alpha e^{-\Gamma t}[(-1 +
e^{(\Gamma -\kappa)t/2}(v + (\kappa-\Gamma)u/\Omega)]\nonumber\\
b_{M+1}(t)&=&-2i\omega\sqrt{r \alpha}e^{-(\Gamma + \kappa) t/2} u/\Omega
\label{eq:cond}
\end{eqnarray} with $\alpha=\gamma_1\gamma/\omega^2$,
$\Omega=\sqrt{4\omega^2-(\kappa-\Gamma)^2}$, $u={\rm sin}[\Omega
t/2]$ and $v={\rm cos}[\Omega t/2]$. An immediate and remarkable
conclusion from this calculation is that the vacuum trapping
condition (i.e. $b_{j=M+1}=0$ but $b_{j\not=M+1}\not=0$) still
arises. Moreover, it will arise for {\em any} number of qubits.
This implies that the effects we have discussed are not just
robust against decoherence: they are to a great extent {\em
immune} to decoherence since $b_{j=M+1}$ is strictly zero at the
renormalized trapping time $\tau^*_c=2m\pi/\Omega$ with $m$ odd,
for any $M$. We now turn to the state fidelity $F_r$ with respect
to the pure system's state at $t=\tau^*$ (see Eq.
\ref{eq:trapstate}), i.e. $F_r=|\langle
\Psi(t=\tau^*)|\widetilde{\Psi}_{cond}(t=\tau_c^*)\rangle|$ where
$|\widetilde{\Psi}_{cond}(t=\tau_c^*)\rangle$ denotes the
normalized conditional state. Several interesting features arises
from the interplay between $\Gamma$ and $\kappa$.  For the
situation in which $\Gamma=\kappa$, the fidelity $F_r$ equals
unity for any value of $r$ and at any time. This is due to the
fact that the non-Hermitian operator accounting for the
dissipative interaction in $\widetilde {H}$, is just the
excitation number-operator (i.e $-i\Gamma \mathcal{N}$) hence the
conditional state becomes $|\Psi_{cond}(t)\rangle=e^{-\Gamma
t}|\Psi (t)\rangle$ and $P(0,t)=e^{-2\Gamma t}$. Therefore the
decoherence sources can effectively be combined to produce a
negligible net effect. This feature becomes more prominent as the
number of qubits increases, as can be seen in Figure
\ref{fig:decoherence}. The state fidelity $F_r$ with $\Gamma \not=
\kappa$, is shown in the two cases in which it is possible to
either generate symmetric W entangled states or to obtain $M$
anti-clones with high fidelity: $r_{W'}$ and $r_{W}^+$. In both
situations the state fidelity moves closer to unity as the number
of qubits increases, since the time interval required to achieve
the desired state becomes shorter. It is also worth noting that
higher values of fidelity are obtained for the symmetric case
$r_{W}^+$ than in the $r_{W'}$ case. This effect can be better
appreciated for a small number of qubits. We note that the
probability of not detecting a photon in $(0,\tau^*_c)$ does not
fall below 0.97 for $M=2$ and becomes even closer to unity for
higher numbers of qubits (see Fig. \ref{fig:decoherence}: inset).
This again shows how efficient our protocols for
entangling/anti-cloning are, and concludes the justification of
the claims in this paper.

In summary, we have shown how asymmetric cavity-qubit
couplings can be exploited to perform very robust, high-fidelity entangling and
anti-cloning operations, in a physically realizable multi-qubit system.

We acknowledge funding from the Clarendon Fund and ORS (AOC), DTI-LINK (NFJ),
and COLCIENCIAS project 1204-05-13614 (LQ).

\end{document}